\documentclass[a4paper,11pt]{article}
\usepackage[utf8]{inputenc}
\usepackage{color}
\usepackage{cite}
\usepackage{hyperref}
\usepackage{amsmath}
\usepackage{graphicx}

\usepackage[top=0.4in, bottom=0.7in, left=0.60in, right=0.5in]{geometry}

\begin{document}
\title{{\Large $(J/\psi, J/\psi)$, and $(\eta_c, \eta_c)$ production through two intermediate photons in electron-positron annihilation at B-factories}}
\author{Shashank Bhatnagar$^1$, Hluf Negash$^2$}
\maketitle \small{1. Department of Physics, University Institute of Sciences, Chandigarh University, Mohali-140413, India\\
\maketitle \small{2. Department of Physics, Aksum University, P.O.Box1010, Axum, Ethiopia\\}

\begin{abstract}
\normalsize{We study the processes, $e^- e^+ \rightarrow \gamma^* \gamma^* \rightarrow J/\psi +J/\psi$, and $e^- e^+ \rightarrow \gamma^* \gamma^* \rightarrow \eta_c+ \eta_c$ at $\sqrt{s}=10.6$ GeV in the framework of $4\times 4$ Bethe-Salpeter equation. For $J/\psi+J/\psi$ production, the dominant contribution is through fragmentation process, while for $\eta_c+\eta_c$ production, the quark rearrangement diagrams contribute. Our results of cross section for $J/\psi+J/\psi$ and $\psi(2S)+\psi(2S)$ are compatible with the experimental upper limits set by Belle Collaboration, while in the absence of experimental data for $\eta_c(1S)+\eta_c(1S)$, and $\eta_c(2S)+\eta_c(2S)$ production, we have given theoretical prediction of their cross sections, and compared with NRQCD prediction.}
\end{abstract}

\section{Introduction}
Double charmonium production in $e^+ e^-$ annihilation at B-factories is one of the most intriguing subjects in quarkonium physics. The experimental measurements of the processes $e^- e+\rightarrow J/\psi +\eta_c$, and $e^- e^+\rightarrow J/\psi+\chi_{cJ}$ had been done by Belle\cite{belle02,belle04} and BABAR\cite{babar05}. However, there was absence of a clear signal for the process $e^+ e^- \rightarrow J/\psi+ J/\psi$ in Belle's measurement\cite{belle04}, though they set an upper limit on the production of $J/\psi+ J/\psi$ at $\sigma(e^+ e^- \rightarrow J/\psi+ J/\psi)\times B( > 2)$ at 9.1 fb \cite{belle04}, where  $B(> 2)$  denotes branching ratio of final states involving more than two charged tracks. The value of cross section for this process was estimated in some works as $\sigma < 9.1$fb \cite{braguta}, by setting $B( > 2) \sim 1$. However, Belle\cite{belle04} ruled out the possibility that in their measurements, a large fraction of the inferred $J/\psi+ \eta_c$ signal consists of $J/\psi +J/\psi$ events. The theoretical studies on double $J/\psi$ production that proceeds through two intermediate photons have been carried out in the framework of NRQCD \cite{braaten03,bodwin03,bodwin06,wang08,huang23}, vector dominance model\cite{fan12}, and the light cone expansion model \cite{braguta}. It is to be noted that through electron-positron annihilation, the states consisting of two charmonia with opposite charge conjugation, such as $J/\psi+\eta_c$ can be produced at order $\alpha^2 {\alpha_s}^2$ through a single virtual photon, whereas the states consisting of charmonium with the same charge conjugation, such as $J/\psi+ J/\psi$, or $\eta_c+\eta_c$ can be produced at order $\alpha^4$ through two virtual photons.

Now, an exclusive quarkonium production process in electron-positron annihilation is dominated by the color-singlet channel, in which a produced $Q\bar{Q}$ pair has the same quantum numbers as that of a final-state quarkonium. Now in  NRQCD, the cross section is factorized into a short distance part (creation of heavy quark pair), and the long distance part (formation of quarkonium), where the short distance component can be calculated perturbatively in strong interaction coupling constant, $\alpha_s$, and is relativistically covariant. However the long distance component is the non-perturbative part, which should incorporate relativistic corrections. Here we wish to mention that for the hard exclusive process, $e^- e^+\rightarrow J/\psi + \eta_c$, there was a large discrepancy between leading order NRQCD\cite{braaten} prediction, and Belle\cite{belle04} data, which could only be resolved when both radiative corrections and relativistic corrections were taken simultaneously.

In the present work, we evaluate the exclusive process, $e^- e^+\rightarrow J/\psi+J/\psi$, and $e^- e^+\rightarrow \eta_c+\eta_c$ in the framework of the Bethe-Salpeter (BSE), which is dynamical equation based approach, that is firmly rooted in quantum field theory, and has a fully relativistic character. It not only incorporates the relativistic effect of quark spins, but can also describe internal motion of constituent quarks within the hadron in a relativistically covariant manner. We wish to point out that it is mainly due to these features, we were able to get substantial contributions to cross section (as  seen from our previous works on: $e^- e^+\rightarrow \gamma + H$ ($H=\chi_{c0}, \chi_{c1}, \eta_c)$\cite{bhatnagar24} at 10.6GeV, and 4.6GeV, and $e^- e^+\rightarrow h_c +\eta_c$, and $e^- e^+\rightarrow h_c +\chi_{c1}$\cite{monika23}) at 10.6GeV, at leading order. Further,  since the energy scale for production processes is large, the perturbative calculation in QCD to lowest order of $\alpha_s$ expansion is expected to be sufficient, since higher order QCD corrections are expected to be negligible. Hence in this work, we  will not consider the contributions from diagrams of higher-order in $\alpha_s$, which is beyond the scope of the paper. Also we do not consider higher order corrections from pure electromagnetic interactions, which should be negligible. Thus in the present work dealing with $(J/\psi, J/\psi)$, and $(\eta_c, \eta_c)$ productions at 10.6 GeV, we will study these processes at leading order in BSE.

Now the $J/\psi+ J/\psi$ production process is governed by two types of diagrams: (A) Photon fragmentation, and (B) Quark rearrangement, to be explained in Section 3. It is to be mentioned that though the production cross section for $J/\psi+ J/\psi$ is suppressed by a kinematical factor, $\frac{\alpha^2}{{\alpha_s}^2}$ relative to processes proceeding through a single virtual photon such as $J/\Psi +\eta_c$ production, but there is enhancement of cross section of the former process due to a kinematical factor associated with the fragmentation diagrams to be discussed in Section 3. Thus, the major contribution to cross section for $J/\psi + J/\psi$ production arises from the fragmentation diagrams, where each of the virtual photons fragments into a $J/\psi$ pair, due to which, some of the authors have considered only the fragmentation contribution\cite{luchinsky,peskin} to cross section for $e^+ e^-\rightarrow J/\psi+ J/\psi$.

Thus in this work, we evaluate this process at the leading order $O(\alpha^4)$, and have calculated the cross section for $e^+ e^- \rightarrow J/\psi+ J/\psi$, and $e^+ e^-\rightarrow \eta_c +\eta_c$ in the framework of $4\times 4$ Bethe-Salpeter equation. We have considered both the fragmentation as well as the quark rearrangement diagrams in $J/\psi+ J/\psi$ production. However the $\eta_c +\eta_c$ production proceeds only through the quark rearrangement diagrams\cite{braaten03}.

The paper is structured as follow: The BSE framework for the $Q\bar{Q}$ system is summarized in Section 2. The cross section calculation for the process, $e^- + e^+ \rightarrow J/\psi + J/\psi$ is covered in Section 3. The calculation for the process, $e^- + e^+ \rightarrow \eta_c + \eta_c$ is covered in Section 4, while Section 5 is devoted to Conclusions.

\section{BSE Framework for Quarkonium bound state}
A 4D Bethe-Salpeter equation (BSE) for quark-antiquark ($Q\bar{Q}$) bound state system can be expressed as,
\begin{equation}
\Psi(P,q)=S_F(p_1)i\int \frac{d^{4}q'}{(2\pi)^{4}}K(q,q')\Psi(P,q')S_F(p_2)
\end{equation}

where the two particles' momenta are $p_1$ and $p_2$, with the hadron's internal momentum being $q$ and the external momentum being $P$ and mass, $M$. In Eq. (1), the interaction kernel is denoted by $K(q,q')$, and the quark and antiquark's inverse propagators are denoted by $S_{F}^{-1}(\pm p_{1,2})=\pm i{\not}p_{1,2}+ m_{1,2}$. The above equation can be reduced to 3D form by applying the Covariant Instantaneous Ansatz (CIA)-a Lorentz-invariant generalization of the Instantaneous Approximation (IA) to the BS kernel, $K(q,q')$, where $K(q,q')=K(\widehat{q},\widehat{q}')$. The BS kernel then depends on the hadron's internal momentum component,  $\widehat{q}_\mu= q_\mu- \frac{q.P}{P^2}P_\mu$ which is a 3-D variable, that is orthogonal to the external hadron momentum, i.e. $\widehat{q}.P=0$, whereas $\sigma P_\mu=\frac{q.P}{P^2}P_\mu$ is the longitudinal component of $q$, and $d^4q=d^3\widehat{q}Md\sigma$ is the 4-D volume element in momentum space. A series of steps detailed in \cite{bhatnagar24} leads to the definition of the 4D BS wave function

\begin{eqnarray}
&&\nonumber \Psi(P, q)=S_1(p_1)\Gamma(\hat q)S_2(-p_2),\\&&
\nonumber \Gamma(\hat{q})=\int \frac{d^3\hat{q}'}{(2\pi)^3}K(\hat{q},\hat{q}')\psi(\hat{q}'),\\&&
\psi(\hat{q}')=i\int\frac{Md\sigma'}{2\pi} \Psi(P,q'),
\end{eqnarray}

with the hadron-quark vertex function, $\Gamma(\hat{q})$ sandwiched between two quark propagators, and is employed for calculation of transition amplitudes for various processes. Also $\psi(\hat{q}')$ is the 3D BS wave function obtained by integrating the 4D BS wave function, $\Psi(P,q')$ over the longitudinal component, $Md\sigma$ of internal hadron momentum as shown in previous equation. Also the 3D reduction of the BSE leads to four Salpeter equations\cite{eshete19}},

\begin{eqnarray}
 &&\nonumber(M-\omega_1-\omega_2)\psi^{++}(\hat{q})=\Lambda_{1}^{+}(\hat{q})\Gamma(\hat{q})\Lambda_{2}^{+}(\hat{q})\\&&
   \nonumber(M+\omega_1+\omega_2)\psi^{--}(\hat{q})=-\Lambda_{1}^{-}(\hat{q})\Gamma(\hat{q})\Lambda_{2}^{-}(\hat{q})\\&&
\nonumber \psi^{+-}(\hat{q})=0.\\&&
 \psi^{-+}(\hat{q})=0\label{fw5}
\end{eqnarray}

It is to be noted that both the 3D Salpeter equations as well as the 4D hadron-quark vertex function depend on the Lorentz-invariant variable, $\hat{q}^2=q^2-(q.P)^2/P^2$\cite{bhatnagar24}, which is a scalar, with validity extending over the entire time-like region of the 4D space, while also keeping contact with the surface, $P.q=0$ (hadron rest frame). The 3D Salpeter equations leads to mass spectral equations that are used not only for the determination of mass spectrum of ground and excited states of various heavy-light mesons\cite{eshete19}, but also the determination of their radial wave functions, that are in turn employed for transition amplitude calculations of various processes, such as leptonic decays, M1 and E1 radiative decays, two-photon decays (for details, see \cite{eshete19,vaishali21, bhatnagar20,vaishali24}). For details of our interaction kernel employed in BSE see \cite{bhatnagar24}. Besides the above low energy transitions, we have also studied the hadron production processes \cite{bhatnagar24, monika23} in $e^- e^+$ annihilation at $\sqrt{s}$= 10.6GeV on lines of \cite{shi20}, where the results of our calculations of these processes were in good agreement with experiment and with other models. Thus so far we have tested our BSE framework for processes over a range of energies from low energy spectroscopy to high energy transitions in an integrated framework, without changing the input assumptions or parameters to suit the needs of different sectors of quarkonium physics.

The details about interaction kernel can be found in \cite{bhatnagar24}. The input parameters employed in our framework are: the spring constant, $\omega_0$ = 0.22 GeV; $A_0=0.01$, and $C_0 = 0.69$, where $\frac{C_0}{\omega_0^2}$ represents the ground state energy. The c-quark's mass is $m_c$ = 1.490 GeV and the QCD length scale is taken as $\Lambda$ = 0.250 GeV (see \cite{bhatnagar24} for details). The input parameters of our model were fitted by solving the corresponding mass spectral equations from our earlier work on calculations mass spectra of heavy-light quarkonia: $0^{++},1^{--}, 0^{-+}, 1^{+-}$, and $1^{++}$ (for more details, see \cite{eshete19}).

\section{Cross section for $e^- e^+\rightarrow J/\psi+ J/\psi$}
In this section, we study the process $e^- e^+ \rightarrow J/\psi +J/\psi$, which proceeds through two virtual photons since  each of the  $J/\psi$ has charge parity, $C=-1$. There are two types of leading-order (LO) Feynman diagrams, shown in Figure 1. In the type 1 process, the quark-anti-quark pair produced by each virtual photon evolves into a vector meson. This is the photon fragmentation diagram, in which the invariant mass of each virtual photon is vector meson mass, $M$. The type 2 process involves electron and positron annihilating to produce two virtual photons, where each virtual photon propagator is of order $\frac{4}{s}$. The two virtual photons create two quark-anti-quark pairs, followed by quark rearrangement, where each produced meson receives quarks from two different virtual photons. Thus due to virtual photon propagators, the photon fragmentation diagrams (Line 1 of Fig.1) are enhanced by a factor $\frac{s^2}{16M^4}\sim 8.588$ in comparison to the quark rearrangement diagrams (in Line 2 of Fig.1) at level of amplitude, $M_{fi}$. The fragmentation contribution is further enhanced by a kinematical factor $log[\frac{\sqrt{s}}{M}]$, originating from collinear emission of virtual photon in the forward direction.

Let $P,q, \epsilon$ and  $P',q', \epsilon'$ be the external momentum, internal momentum and the polarization vectors of the two outgoing vector mesons. We take $k$ and $k'$ as the momenta of the two virtual photons involved in the process.

\begin{figure}[ht!]
 \centering
 \includegraphics[width=14cm,height=8cm]{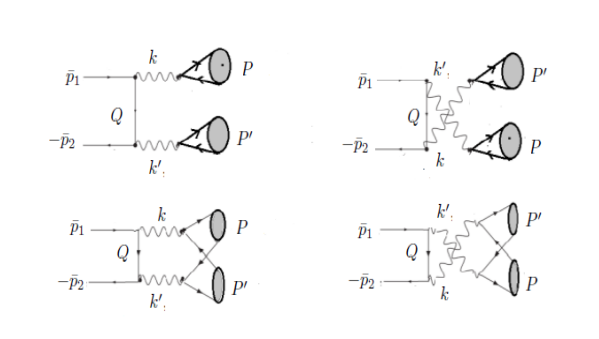}
 \caption{Feynman diagrams involved in $e^- e^+ \rightarrow J/\psi+ J/\psi$ at leading order. Line 1 comprises of the photon fragmentation diagrams, while line 2 comprises of the quark rearrangement diagrams.}
 \end{figure}

For the photon fragmentation diagrams (in line 1 of Fig.1), one can write the amplitude as,

\begin{equation}
{M^1}_{fi}|_D=\frac{e^2e_Q^2}{M^2M'^2}[\bar{v}^{(s2)}(\bar{p_2})ie\gamma_{\nu}[\frac{-i{\not}Q+m_e}{Q^2+m_e^2}]ie\gamma_{\mu}u^{(s1)}(\bar{p_1})]\int \frac{d^4q}{(2\pi)^4}Tr[ie_Q\gamma_{\mu}\bar{\Psi}(P,q)] \int \frac{d^4q}{(2\pi)^4}Tr[ie_Q\gamma_{\nu}\bar{\Psi}(P',q')]
\end{equation}

The amplitude for the other fragmentation diagram that involves exchange of final state mesons for this process can again be obtained from the above amplitude through the exchange of indices as, $M^2_{fi}|_X=M^2_{fi}|_{D}~{\mu \leftrightarrow \nu}$. Thus,

\begin{equation}
{M^1}_{fi}|_X=\frac{e^2e_Q^2}{M^2M'^2}[\bar{v}^{(s2)}(\bar{p_2})ie\gamma_{\nu}[\frac{-i{\not}Q+m_e}{Q^2+m_e^2}]ie\gamma_{\mu}u^{(s1)}(\bar{p_1})]\int \frac{d^4q}{(2\pi)^4}Tr[ie_Q\gamma_{\nu}\bar{\Psi}(P,q)] \int \frac{d^4q}{(2\pi)^4}Tr[ie_Q\gamma_{\mu}\bar{\Psi}(P',q')].
\end{equation}

To reduce the above amplitudes to 3D form, we make use of the Covariant Instantaneous Ansatz (CIA), where the 4D volume elements of internal momenta of the two produced hadrons can be written as $d^4q=d^3\hat{q} Md\sigma$ (and $d^4q'=d^3\hat{q}' M'd\sigma'$). We then integrate over the longitudinal components $Md\sigma$, and $M'd\sigma'$ of internal momenta $q$ and $q'$ of the two hadrons to obtain,$\int \frac{Md\sigma}{2\pi i}\hat{\Psi}(P,q)=\bar{\Psi}(\hat{q})$, and $\int \frac{M'd\sigma'}{2\pi i}\hat{\Psi}(P',q')=\bar{\Psi}(\hat{q}')$.

Thus, for the Direct amplitude of photon fragmentation diagrams, we reduce the amplitude to 3D form as,
\begin{equation}
{M^1}_{fi}|_D=\frac{e^2e_Q^2}{M^2M'^2}[\bar{v}^{(s2)}(\bar{p_2})\gamma_{\nu}[\frac{-i{\not}Q+m_e}{Q^2+m_e^2}]\gamma_{\mu}u^{(s1)}(\bar{p_1})]\int \frac{d^3\hat{q}}{(2\pi)^3}Tr[\gamma_{\mu}\bar{\Psi}(\hat{q})] \int \frac{d^3\hat{q}}{(2\pi)^3}Tr[\gamma_{\nu}\bar{\Psi}(\hat{q}')]
\end{equation}

The amplitude for the exchange diagram for this process can be then be similarly written as,

\begin{equation}
{M^1}_{fi}|_X=\frac{e^2e_Q^2}{M^2M'^2}[\bar{v}^{(s2)}(\bar{p_2})\gamma_{\nu}[\frac{-i{\not}Q+m_e}{Q^2+m_e^2}]\gamma_{\mu}u^{(s1)}(\bar{p_1})]\int \frac{d^3\hat{q}}{(2\pi)^3}Tr[\gamma_{\nu}\bar{\Psi}(\hat{q})] \int \frac{d^3\hat{q}}{(2\pi)^3}Tr[\gamma_{\mu}\bar{\Psi}(\hat{q}')].
\end{equation}

\bigskip
\bigskip

Similarly at leading order, the amplitude, $M^{2}_{fi}$ for the quark rearrangement diagrams (in Line $\#2$ of Fig. 1 ) is expressed in lowest order as,
\begin{equation}
 M^{2}_{fi}|_D=\frac{16e^2{e_Q}^2}{s^2}[\bar{v}^{(s2)}(\bar{p_2})ie\gamma_{\nu}[\frac{-i{\not}Q+m_e}{Q^2+m_e^2}]ie\gamma_{\mu}u^{(s1)}(\bar{p_1})]\int \frac{d^4q}{(2\pi)^4}\int \frac{d^4q}{(2\pi)^4} Tr[ie_Q\gamma_{\mu}\bar{\Psi}(P',q')(ie_Q\gamma_{\nu})\bar{\Psi}(P,q)],
\end{equation}

and the amplitude  for the exchange diagram of quark rearrangement process, above can be obtained by exchange of covariant indices: ${M^2}_{fi}|_X={M^2}_{fi}|_{D}  (\mu \leftrightarrow \nu)$, that is characterized by exchange of two final state vector mesons, the amplitude is expressed as:

\begin{equation}
 {M^2}_{fi}|_X=\frac{16e^2{e_Q}^2}{s^2}[\bar{v}^{(s2)}(\bar{p_2})ie\gamma_{\nu}[\frac{-i{\not}Q+m_e}{Q^2+m_e^2}]ie\gamma_{\mu}u^{(s1)}(\bar{p_1})]\int \frac{d^4q}{(2\pi)^4}\int \frac{d^4q}{(2\pi)^4} Tr[ie_Q\gamma_{\nu}\bar{\Psi}(P',q')(ie_Q\gamma_{\mu})\bar{\Psi}(P,q)].
\end{equation}

Here, in both the types of diagrams, the 4D adjoint BS wave functions can be written as,

\begin{eqnarray}
&&\nonumber \bar{\Psi}_{V}(P, q)=S_F(-p_2)\Gamma_V(\hat{q})S_F(p_1)\\&&
\bar{\Psi}_V(P', q')=S_F(-p_2)\Gamma_V(\hat{q}')S_F(p_1)
\end{eqnarray}

Here, we take into account $p_{1,2}$ as the momenta of quarks for one of the produced hadrons with total momentum $P$, and internal momentum, $q$, while $p'_{1,2}$ are the momenta of the two quarks of the other produced hadron with momentum $P'$, and internal momentum, $q'$. The quark momenta are expressed in terms of the internal and total momenta of the corresponding mesons as follows:

\begin{equation}
p_1=\frac{1}{2}P + q,~~~~ p_2=\frac{1}{2}P -q,~~~~p'_1 =\frac{1}{2}P' -q',~~~~~~~ p'_2=\frac{1}{2}P'+q'.
\end{equation}

Both the incoming and outgoing particle's wave functions are normalised to one particle per unit volume. The expressions above may be factored into the product of leptonic tensor and hadronic tensor: $M^1_{fi}=L_{\mu \nu}^{em} H_{\mu \nu}^{hadronic}$. The reduction of amplitude for the quark rearrangement diagram (on left of Line $\#2$ of Fig.1) to 3D form as explained above, leads to:

\begin{equation}
 M^2_{fi}|_D=\frac{16e^2{e_Q}^2}{s^2}[\bar{v}^{(s2)}(\bar{p_2})ie\gamma_{\nu}[\frac{-i{\not}Q+m_e}{Q^2+m_e^2}]ie\gamma_{\mu}u^{(s1)}(\bar{p_1})]\int \frac{d^3\hat{q}}{(2\pi)^4}\int \frac{d^3\hat{q}}{(2\pi)^4} Tr[ie_Q\gamma_{\mu}\bar{\Psi}(\hat{q}')(ie_Q\gamma_{\nu})\bar{\Psi}(\hat{q})].
\end{equation}

Similarly for exchange diagram (on right of Line $\#2$ of Fig. 1) for this process,
\begin{equation}
 M^2_{fi}|_X=\frac{16e^2{e_Q}^2}{s^2}[\bar{v}^{(s2)}(\bar{p_2})ie\gamma_{\nu}[\frac{-i{\not}Q+m_e}{Q^2+m_e^2}]ie\gamma_{\mu}u^{(s1)}(\bar{p_1})]\int \frac{d^3\hat{q}}{(2\pi)^4}\int \frac{d^3\hat{q}}{(2\pi)^4} Tr[ie_Q\gamma_{\nu}\bar{\Psi}(\hat{q}')(ie_Q\gamma_{\mu})\bar{\Psi}(\hat{q})].
\end{equation}

Now, total amplitude, $M_{fi}=M^{1}_{fi} +M^{2}_{fi}$, where, $M^{1}_{fi}={M^{1}_{fi}}_{D} +{M^{1}_{fi}}_{X}$, and $M^{2}_{fi}={M^{2}_{fi}}_{D} +{M^{2}_{fi}}_{X}$. However, in order to calculate $M_{fi}$ further, one needs the algebraic forms of 3D adjoint BS wave functions, $\bar{\Psi}_{V}(\hat{q})$, of vector mesons. For this, we first start with the most general form of 4D BS wave function, $\Psi_P(P,q)$ for vector meson $1^{--}$, that is expressed in terms of various Dirac structures in \cite{smith69,alkofer02}. Then, making use of the 3D reduction under Covariant Instantaneous Ansatz, and making use of the fact that $\hat{q}.P=0$, we can write the general decomposition of the instantaneous BS wave function for vector meson of dimensionality $M$ being composed of various Dirac structures that are multiplied with scalar functions $\phi_i(\hat{q})$ as in \cite{wang22}. The amplitudes $\phi_1(\hat{q})$ are all independent. Putting this wave function into the Salpeter equations, Eq.(6) leads to the 3D wave function, $\Psi_V(\hat{q})$, whose adjoint wave function, $\bar{\Psi}_V(\hat{q})$ is obtained as, $\bar{\Psi}_V(\hat{q})=\gamma_4 \Psi^\dag_V(\hat{q})\gamma_4$:

\begin{equation}
\bar{\Psi}_V(\hat{q})=N_V\bigg[iM{\not}\epsilon+(\hat{q}.\epsilon)\frac{M(m_1+m_2)}{\omega_1\omega_2+m_1m_2-\hat{q}^2}+{\not}{P}{\not}{\epsilon}+i\frac{\omega_1+\omega_2}{2(\omega_1m_2+m_1\omega_2)}({\not}P{\not}\epsilon {\not}\hat{q}+\hat{q}.\epsilon)\bigg]\phi_V(\hat{q})
\end{equation}

which for the two equal mass mesons reduces to,

\begin{eqnarray}
&&\nonumber \bar{\Psi}_V(\hat{q})=N_V\bigg[iM{\not}\epsilon+(\hat{q}.\epsilon)\frac{M}{m}+{\not}{P}{\not}{\epsilon}+\frac{i}{2m}{\not}\hat{q}{\not}\epsilon{\not}P+\frac{i}{2m}(\hat{q}.\epsilon){\not}P\bigg]\phi_V(\hat{q}),\\&&
\bar{\Psi}_V(\hat{q}')=N'_V\bigg[iM'{\not}\epsilon'+(\hat{q}'.\epsilon')\frac{M'}{m}+{\not}{P}'{\not}{\epsilon}'+\frac{i}{2m}{\not}\hat{q}'{\not}\epsilon'{\not}P'+\frac{i}{2m}(\hat{q}'.\epsilon'){\not}P'\bigg]\phi_V(\hat{q}').
\end{eqnarray}

where $N_V$ is the BS normalizer evaluated through the current conservation condition to be discussed later. Also, the 3D radial wave function $\phi_V(\hat{q})$ satisfies the mass spectral equation that resembles the equation of a 3D harmonic oscillator (see Refs.\cite{eshete19} for details). Expressing the laplacian operator in this equation in spherical polar co-ordinates, we express mass spectral equation as a  partial differential equation in $\phi_V(\hat{q})$ in terms of orbital angular momentum quantum number, $l = 0,1,2…$ corresponding to $S, P, D,...$ wave states. The solutions of this 3D harmonic oscillator equation is obtained using a power series method, where the eigen values of this equation are: $E_N=2\beta^2 (N +3/2)$, where the principal quantum number, $N=2n+l$; with $n= 0,1,2,…$. We take $l=0 (S$ state) and $l=2 (D$ state)  for vector ($1^{--}$) mesons. Thus, solutions of mass spectral equation\cite{eshete19} not only leads to spectrum of quarkonia, but also their radial wave functions that are wave functions with definite quantum numbers, $N$ and $l$.

The analytic structures of the 3D wave functions \cite{bhatnagar20} for $1^{--}$ mesons are\cite{eshete19}:
\begin{eqnarray}\label{wavefunc}
&&\nonumber \phi_V(1S,\hat q)= e^{-\frac{\hat{q}^2}{2{\beta_V}^2}};\\&&
\phi_V(2S,\hat{q})=(1-\frac{2\hat{q}^2}{3{\beta_V}^2})e^{-\frac{\hat{q}^2}{2{\beta_V}^2}},
\end{eqnarray}

where $\beta_V=(Mm{\omega_0}^2 \alpha_s)^{1/4}$ (with $\alpha_s(M^2)=\frac{12\pi}{33-2f}(log\frac{M^2}{\Lambda^2})^{-1}$ being the QCD coupling constant) is the inverse range parameter\cite{eshete19} for vector meson. The analytic structure of 3D wave functions, $\phi_V(nS, \hat{q}')$ of the second produced vector meson can be obtained with the replacement, $\hat{q} \rightarrow \hat{q}'$ in the above equations. After evaluation of the trace over $\gamma$ matrices (where, $\vec{\bar{p_1}}=- \vec{\bar{p_2}}$, and $E_1=E_2 (=E)$), $\sqrt{s}=2E$, the amplitude, ${M^1}_{fi}$ for photon fragmentation diagrams (with contributions from both the Direct and the Exchange diagrams) is then written in the centre of mass frame as,

\begin{eqnarray}
&&\nonumber M^{1}_{fi}=\frac{e^2{e_Q}^2}{M^2 M'^2}\int \frac{d^3\hat{q}'}{(2\pi)^3}\phi_V(\hat{q}') \int \frac{d^3\hat{q}}{(2\pi)^3}\phi_V(\hat{q})\\&&
\nonumber \bigg[(16MM') \{[\bar{v}^{(s2)}(\bar{p_2}){\not}\epsilon\frac{(-i{\not}Q+m_e)}{Q^2+m_e^2}{\not}\epsilon' u^{(s1)}(\bar{p_1})]+[\bar{v}^{(s2)}(\bar{p_2}){\not}\epsilon'\frac{(-i{\not}Q+m_e)}{Q^2+m_e^2}{\not}\epsilon u^{s1}(\bar{p_1})] \}+\\&&
\nonumber 16\frac{M'}{m}(\hat{q}.\epsilon)\{[\bar{v}^{(s2)}(\bar{p_2}){\not}P\frac{(-i{\not}Q+m_e)}{Q^2+m_e^2}{\not}\epsilon'u^{s1}(\bar{p_1})]+ [\bar{v}^{(s2)}(\bar{p_2}){\not}\epsilon'\frac{(-i{\not}Q+m_e)}{Q^2+m_e^2}{\not}P u^{(s1)}(\bar{p_1})]\}+\\&&
\nonumber 16\frac{M'}{m}(\hat{q}'.\epsilon')\{[\bar{v}^{(s2)}(\bar{p_2}){\not}\epsilon\frac{(-i{\not}Q+m_e)}{Q^2+m_e^2}{\not}P'u^{s1}(\bar{p_1})]+ [\bar{v}^{(s2)}(\bar{p_2}){\not}P'\frac{(-i{\not}Q+m_e)}{Q^2+m_e^2}{\not}\epsilon u^{(s1)}(\bar{p_1})]\}+\\&&
\frac{16}{m^2}(\hat{q}.\epsilon)(\hat{q}'.\epsilon')\{[\bar{v}^{(s2)}(\bar{p_2}){\not}P\frac{(-i{\not}Q+m_e)}{Q^2+m_e^2}{\not}P']u^{s1}(\bar{p_1})+ [\bar{v}^{(s2)}(\bar{p_2}){\not}P'\frac{(-i{\not}Q+m_e)}{Q^2+m_e^2}{\not}P u^{(s1)}(\bar{p_1})]\}\bigg].
\end{eqnarray}

Similarly for quark rearrangement diagrams we can write the amplitude of the process as,

\begin{eqnarray}
&&\nonumber M^{2}_{fi}|=(e^2e_Q^2) \frac{16}{s^2}\int \frac{d^3\hat{q}'}{(2\pi)^3}\phi_V(\hat{q}') \int \frac{d^3\hat{q}}{(2\pi)^3}\phi_V(\hat{q})\\&&
\nonumber \bigg[(-4MM'+4P.P'+\frac{\hat{q}'.\hat{q} P'.P}{m^2})[\bar{v}^{(s2)}(\bar{p_2}){\not}\epsilon\frac{(-i{\not}Q+m_e)}{Q^2+m_e^2}{\not}\epsilon' u^{(s1)}(\bar{p_1})+\bar{v}^{(s2)}(\bar{p_2}){\not}\epsilon'\frac{(-i{\not}Q+m_e)}{Q^2+m_e^2}{\not}\epsilon u^{(s1)}(\bar{p_1})]+\\&&
\nonumber (-\frac{4MM'(\hat{q}.\epsilon)(\hat{q}'.\epsilon')}{m^2}+4MM'(\epsilon.\epsilon'))[\bar{v}^{(s2)}(\bar{p_2})\frac{(-i2{\not}Q+4m_e)}{Q^2+m_e^2}{u}^{(s1)}(\bar{p_1})]+\\&&
\nonumber 2[\frac{(\hat{q}.\hat{q}')(\epsilon.\epsilon')}{m^2}-\frac{(\hat{q}.\epsilon')(\hat{q}'.\epsilon)}{m^2}][\bar{v}^{(s2)}(\bar{p_2}){\not}P\frac{(-i{\not}Q+m_e)}{Q^2+m_e^2}{\not}P'{u}^{(s1)}(\bar{p_1})+\bar{v}^{(s2)}(\bar{p_2}){\not}P'\frac{(-i{\not}Q+m_e)}{Q^2+m_{e}^2}{\not}P{u}^{(s1)}(\bar{p_1})]\\&&
\nonumber \frac{2M}{m}(\hat{q}'.\epsilon)[\bar{v}^{(s2)}(\bar{p_2}){\not}P'\frac{(-i{\not}Q+m_e)}{Q^2+m_e^2}{\not}\epsilon'{u}^{(s1)}(\bar{p_1})+\bar{v}^{(s2)}(\bar{p_2}){\not}\epsilon'\frac{(-i{\not}Q+m_e)}{Q^2+m_{e}^2}{\not}P'{u}^{(s1)}(\bar{p_1})]+\\&&
\nonumber \frac{4M}{m}(\hat{q}'.\epsilon')[\bar{v}^{(s2)}(\bar{p_2}){\not}P'\frac{(-i{\not}Q+m_e)}{Q^2+m_e^2}{\not}\epsilon {u}^{(s1)}(\bar{p_1})+\bar{v}^{(s2)}(\bar{p_2}){\not}\epsilon\frac{(-i{\not}Q+m_e)}{Q^2+m_{e}^2}{\not}P'{u}^{(s1)}(\bar{p_1})]-\\&&
\nonumber \frac{(P'.P)(\hat{q}'.\epsilon)}{m^2}[\bar{v}^{(s2)}(\bar{p_2}){\not}\hat{q}\frac{(-i{\not}Q+m_e)}{Q^2+m_e^2}{\not}\epsilon' {u}^{(s1)}(\bar{p_1})+\bar{v}^{(s2)}(\bar{p_2}){\not}\epsilon'\frac{(-i{\not}Q+m_e)}{Q^2+m_{e}^2}{\not}\hat{q}{u}^{(s1)}(\bar{p_1})]-\\&&
\nonumber \frac{4M'(\hat{q}.\epsilon)}{m}[\bar{v}^{(s2)}(\bar{p_2}){\not}\epsilon'\frac{(-i{\not}Q+m_e)}{Q^2+m_e^2}{\not}P {u}^{(s1)}(\bar{p_1})+\bar{v}^{(s2)}(\bar{p_2}){\not}P\frac{(-i{\not}Q+m_e)}{Q^2+m_{e}^2}{\not}\epsilon'{u}^{(s1)}(\bar{p_1})]-\\&&
 \frac{(P'.P)(\hat{q}.\epsilon')}{m^2}[\bar{v}^{(s2)}(\bar{p_2}){\not}\hat{q}\frac{(-i{\not}Q+m_e)}{Q^2+m_e^2}{\not}\hat{q}' {u}^{(s1)}(\bar{p_1})+\bar{v}^{(s2)}(\bar{p_2}){\not}\hat{q}'\frac{(-i{\not}Q+m_e)}{Q^2+m_{e}^2}{\not}\hat{q}{u}^{(s1)}(\bar{p_1})]\bigg]
\end{eqnarray}

The above two expressions can be further compactified by introducing $G_1, G'_1$ (for vector meson $\#$ 1), and $G_2, G'_2$ (for vector meson $\#$ 2) as the expressions for the 3D integrals involving the two outgoing vector mesons, defined as:
\begin{eqnarray}
&&\nonumber G_1=\int \frac{d^3\hat{q}}{(2\pi)^3}\phi_V(\hat{q})\\&&
\nonumber G'_1= \int \frac{d^3\hat{q}}{(2\pi)^3}|\hat{q}|\phi_V(\hat{q})\\&&
\nonumber G_2=\int \frac{d^3\hat{q}'}{(2\pi)^3}\phi_V(\hat{q}')\\&&
G'_2=\int \frac{d^3\hat{q}'}{(2\pi)^3}|\hat{q}'|\phi_V(\hat{q}')
\end{eqnarray}

Here, we have made use of the fact that $\hat{q}$ is an effective 3D variable, due to which we can write $\hat{q}.\epsilon=\hat{q}_{\nu}\epsilon_{\nu}=|\hat{q}|(I.\epsilon)$, where, $I_{\mu}$ is a unit vector along the direction of $\hat{q}_{\nu}$, and is expressed as, $I=\frac{\hat{q}}{|\hat{q}|}$, where $|\hat{q}|=\sqrt{\hat{q}^2}=\sqrt{q^2-(q.P)^2/P^2}$, and is a Lorentz-invariant variable \cite{bhatnagar20,wang22}. Here $P'.\epsilon=0$, due to the fact that $P.\epsilon=\overrightarrow{P}.\overrightarrow{\epsilon}=0$, since momentum, $P=(\overrightarrow{P}, iE)$ of vector meson is always orthogonal to its polarization vector, $\epsilon=(\overrightarrow{\epsilon},i0)$. And in center of mass frame, $\overrightarrow{P}'=-\overrightarrow{P}$, which implies that $\overrightarrow{P}'.\overrightarrow{\epsilon}=P'.\epsilon=0$.

The total amplitude for the process will receive contributions from both the fragmentation and the rearrangement diagrams, and can be written as:
\begin{equation}
M_{fi}=M^{1}_{fi} + M^{2}_{fi}.
\end{equation}

However, it is seen that the term on the left and right in every line of the amplitudes in Eqs.(17) and (18) contribute equally.  We now evaluate the spin averaged invariant amplitude modulus squared, $|\bar{M}_{fi}|^2=\frac{1}{4}\sum_{s1,s2,\lambda,\lambda'}{M^\dag}_{fi}M_{fi}$, for which we must now average over the initial spin states of $e^-$ and $e^+$ and sum over the polarisation states of the two final vector mesons. We make use of
$\Sigma_{\lambda} \epsilon_{\mu}^{\lambda} \epsilon_{\nu}^{\lambda} = \frac{1}{3}(\delta_{\mu\nu} + \frac{P_{\mu}P_{\nu}}{M^2})$ for the polarisation vectors of the vector meson. The  electromagnetic coupling constant ($\alpha_{em}=\frac{e^2}{4\pi}$), $m$ is the charm quark mass and $m_e$  is the electron mass.

The kinematics employed  in the centre of mass frame is as follows: Taking $\theta$ to be the angle between the incident beam direction and $J/\Psi$, the dot products of various momenta can be expressed as: $(\bar{p_1}+\bar{p_2})^2=-s$, ~$\bar{p_1}.\bar{p_2}=-\frac{s}{2}$,~ $Q^2=\frac{s}{4}$, ~$\bar{p}_1.P=-\frac{s}{4}(1+Cos\theta)=\bar{p}_2.P'$, ~ $\bar{p}_1.P'=\frac{s}{4}(-1+Cos\theta)=\bar{p}_2.P$,~~$Q.P=\frac{s}{4}Cos\theta$, ~~ $Q.P'=-\frac{s}{4}Cos\theta$. Similarly,
$\bar{p_1}.Q=-\bar{p_2}.Q  \sim \frac{s}{4}$,~ ~$P.P'=-\frac{s}{2}+M^2$,~~ $P.I=0$,~~ $P'.I=0$. And $P.\epsilon=P'.\epsilon'=0$,~ ~ $P.\epsilon'=P'.\epsilon=0$. Similarly, the dot products: $p_1.I=p_2.I'=\frac{\sqrt{s}}{2}Sin\theta$,~~ $p_1.I'=p_2.I=-\frac{\sqrt{s}}{2}Sin\theta$, ~~ $I.Q=\frac{\sqrt{s}}{2}Sin\theta$, and $I'.Q=-\frac{\sqrt{s}}{2}Sin\theta$.

We do not give here the expression for $|\bar{M}_{fi}|^2$, which is quite lengthy and involved, and present here the individual expressions, $|\bar{M}^{1}_{fi}|^2$ for fragmentation diagrams, and $|\bar{M}^{2}_{fi}|^2$ for quark rearrangement diagrams. The former can be expressed as:

\begin{equation}
 |\bar{M}^{1}_{fi}|^2= N_{V}^2 N_{V'}^2[\Omega+\Omega'],
\end{equation}

where, $[\Omega]=\Sigma_{i} [TR]_{ii}$, and $[\Omega']=\Sigma_{i\neq j}[TR]_{ij}$, where $[TR]$ is the trace over gamma matrices in evaluation of $ |\bar{M}^{1}_{fi}|^2$.  The final expression for $[\Omega]$ (taking equal masses ($M'=M$), and the 3D integrals, $G_2=G_1$,~ $G'_2=G'_1$) of identical final state mesons is written as:

\begin{eqnarray}
&&\nonumber [\Omega] = \frac{2^{24}\pi^4\alpha^4}{3^4 m^4 M^8(4m_e^{2}+s)^2}\bigg[F_1+F_2 Cos\theta+F_3 Cos^2\theta+F_4 s^3Cos^3\theta\bigg];\\&&
\nonumber F_1=2{G_{1}}^{2}{G'_{1}}^{2}m^2M^2(4M^2 s^2+7s^3)+2G_{1}^{4}m^4(8M^4s^2+s^4)+{G'_{1}}^{4}(4M^{4} s^2-5M^2 s^3+s^4)\\&&
\nonumber F_2=-4({G'_{1}}^{4}+{G_{1}}^{4}m^4)s^4+4M^2[13{G_{1}}^{2}{G'_{1}}^{2}m^{2} s^3+6{G_{1}}^{4}m^4 s^3+2{G'_{1}}^{4}(-2M^{2} s^2+s^3)]\\&&
\nonumber F_3=-2{G_{1}}^{2}{G'_{1}}^{2} m^2(29 M^2 s^3-2 s^4)+4{G'_{1}}^{4}(5M^2 s^3-s^4)+{G_{1}}^{4} m^4(-24M^2 s^3+s^4)\\&&
F_4={G'_{1}}^4(8M^2-4s)+4{G_{1}}^2 {G'_{1}}^{2} m^2 s-2{G_{1}}^{4}m^4 s+{G_{1}}^{4} m^4 sCos\theta
\end{eqnarray}

while the expression for $[\Omega']$ that involves the cross terms in the trace calculation is:

\begin{eqnarray}
&&\nonumber [\Omega']=\frac{2^{21}\pi^4 \alpha^4 G_1G'_1}{3^4m^3M^8(4m_{e}^2+s)^2}\bigg[H_1+32s^{7/2}Cos\frac{\theta}{2}Sin^3\frac{\theta}{2}H_2+H_3Cos\theta+H_4Cos2\theta+\\&&
\nonumber ~~~~~~~~~~~~~~~~~~~~~~~~~~~~~~~~~~~~~~~~~~~~~~~~~~~~~~~~~H_5Cos3\theta+H_6Cos4\theta+H_7s^{5/2}Sin\theta+H_8 s^{5/2}Sin2\theta\bigg];\\&&
\nonumber H_1=G_{1}G'_{1}m M^{2}(125s^{3}-160M^{2}s^2),\\&&
\nonumber H_2={G_{1}}^{2}m^{3}-3 {G'_{1}}^{2}M+(2{G'_{1}}^{2}+{G_{1}}^{2}m^{2})MCos\theta+{G_{1}}^{2}m^{3}Cos2\theta,\\&&
\nonumber H_3=24M^{2} G_{1} G'_{1}m s^{3},\\&&
\nonumber H_4=2M^{2} G_{1}G'_{2}m(16 M^{2}s^{2}+23s^{3}),\\&&
\nonumber H_5=4M^{2} G_{1}G'_{1}m s^{3},\\&&
\nonumber H_6=M^{2}G_{1} G'_{1} m s^{3},\\&&
\nonumber H_7=-16M^{2}[2{G_{1}}^{2}m^{2}(4m-3M)+7{G'_{1}}^{2} M],\\&&
H_8=8M^{2}[-{G'_{1}}^2M+4{G_{1}}^{2}m^{2}(4m+M)]
\end{eqnarray}
For quark rearrangement diagrams, $|M^{2}_{fi}|^2$ simplifies  when taking $M'=M,~ G_2=G_1$ and $G'_2=G'_1$ for identical final states of the same mass, and ignoring the contributions from the sub-leading order Dirac structures\cite{vaishali21} in $\Psi_V(\hat{q})$, that should have negligible contribution.  We write
\begin{equation}
 |\bar{M}^{2}_{fi}|^2= N_V^2 N_{V'}^2[\Sigma+\Sigma'],
\end{equation}
 where the expressions for $[\Sigma]$ and $[\Sigma']$ (defined in a similar manner to $[\Omega]$, and $[\Omega']$ above) are:

\begin{eqnarray}
&&\nonumber [\Sigma]=\frac{2^{16}\times 131}{3^6}\frac{\alpha^4\pi^4}{m^4M^4s^4(4m_{e}^2+s)^2}\bigg[Y_1+Y_2Cos\theta+Y_3Cos^2\theta+Y_4Cos^3\theta+Y_5Cos^4\theta\bigg]\\&&
\nonumber Y_1=G_{1}^4m^4[2s^6576M^8s^2-720M^6s^3+144M^4s^4]\\&&
\nonumber Y_2=144G_{1}^4m^4M^4(4M^4s^2-4M^2s^3+s^4)\\&&
\nonumber Y_3=-2G_{1}^4m^4[-72M^4s^4+360M^6s^3-8M^2s^5]\\&&
\nonumber Y_4=144G_{1}^4m^4M^4s^3(2M^2-s)\\&&
Y_5=2G_{1}^2m^4s^6
\end{eqnarray}

\begin{eqnarray}
&&\nonumber [\Sigma']=\frac{2^{21}}{3^6}\frac{\alpha^4\pi^4 G_{1}^2}{m^2M^4s^4(4m_{e}^2+s)^2}\bigg[Z_1+Z_2Cos\theta+Z_3Cos2\theta+Z_4Cos3\theta+Z_5Cos4\theta\bigg]\\&&
\nonumber Z_1=-3616{G'_{1}}^{2}M^{6} s^{3}+7840{G_{1}}^2 m^2M^6s^3+2176{G'_{1}}^2 M^6 s^3-7264{G_{1}}^2 m^{2} M^{6} s^3-7836{G_{1}}^ 2m^{2}M^{4} s^{4}+\\&&
\nonumber ~~~~~~~~~~~ 5392G_{1}^2m^2M^4s^4+764G_{1}^2m^2M^2s^5-202G_{1}^2m^2s^6\\&&
\nonumber Z_2=s^3[448{G'_{1}}^2 M^6+{G_{1}}^2  m^2s(944M^4-254M^{2} s+87s^2)]\\&&
\nonumber Z_3=1184{G'_{1}}^2 M^6 s^3-2{G_{1}}^2 m^2(163s^3-192M^{6}s^3+556 M^{4} s^4-426s^5)\\&&
\nonumber Z_4=48G_{1}^2m^2M^4s^4-18{G_{1}}^2 m^2 M^2 s^5+5{G_{1}}^2 m^2 s^6\\&&
Z_5=-36G_{1}^2(m^{2} M^{4} s^{4}+m^{2} M^{2} s^{5}-8m^{2} s^{6})
\end{eqnarray}

Further, $N_V$ is the 4D BS normalizer of each of the identical vector mesons, obtained through the current conservation condition,

\begin{equation}\label{46}
2iP_\mu=\int \frac{d^{4}q}{(2\pi)^{4}}
\mbox{Tr}\left\{\overline{\Psi}(P,q)\left[\frac{\partial}{\partial
P_\mu}S_{F}^{-1}(p_1)\right]\Psi(P,q)S_{F}^{-1}(-p_2)\right\} +
(1\rightleftharpoons2).
\end{equation}

In terms of the transition rate per unit volume, $W_{fi}$, we may express $d\sigma$ as follows:
\begin{equation}
d\sigma=\frac{W_{fi}(2\pi)^4\delta^{(4)}(p_1+p_2-P'-P'')}{F}[\frac{d^3\vec{P}'}{2E'(2\pi)^3}\frac{d^3\vec{P}''}{2E''(2\pi)^3}].
\end{equation}

In center-of-mass frame, the incident flux, $F=4\sqrt{(\bar{p_1}.\bar{p_2})^2-m_1^2 m_2^2}$  reduces to $F=4|\vec{p}|\sqrt{s}$, where $|\vec{p}|$ is the magnitude of momentum of either of the incident particles, and $\sqrt{s}=2E$ is the centre of mass energy of the incident particles. The entire cross section following a series of steps can be expressed as:

\begin{equation}
\sigma=\frac{1}{32\pi^2 s^{3/2}}|\vec{P'}|\int d\Omega' |\bar{M}_{fi}|^2,
\end{equation}

with $|\vec{P'}|=\sqrt{\frac{1}{s}[s-(M+M')^2][s-(M-M')^2]}$ the momenta of either of the ejecting particles. It is important to note that for $V V$ (vector-vector) final states involving 2 identical particles, the angular distribution must be integrated over only one hemisphere, while for $V1V2 (V1 \neq V2)$, the integration should be done over the full angular distribution. Now, it is possible to verify that the BS normalizers behave as $N_V \sim M^{-2}$, and $N'_V\sim M^{-2}$. The 3D integrals behave as: $G_1\sim M^3,~G'_1\sim M^4$. Consequently, $|\bar{M}_{fi}|^2 \sim M^0$. As a result, the cross sectional formula exhibits the behaviour $\sigma \sim M^{-2}$.

Thus the amplitude and the cross section for the production process, $e^+ e^-\rightarrow J/\psi+J/\psi$ is represented in terms of the radial wave functions $\phi_V(\hat{q})$, of both the produced quarkonia, which were obtained analytically from solutions of mass spectral equations of vector mesons\cite{eshete19}. The cross sections for $e^- e^+\rightarrow J/\psi J/\psi$, with $n,n'=1,2$ calculated in this work are given in Table 2 along with results of other models\cite{braaten03,luchinsky,braguta}, though the experimental results of this process as mentioned in the Introduction, are not yet available. It is seen that the integrals, $G_1, G_2$ (where $G_1=G_2$), and $G'_1, G'_2$ (where $G'_1=G'_2$ for two identical vector mesons in final state) have a major role to play in the calculation of the cross section. Their numerical values along with the values of BS normalizers $N_V$, for ground and excited states of V mesons are listed in Table 2.

\begin{table}[hhhhh]
  \begin{center}
  \begin{tabular}{p{5.3cm} p{2.2cm} p{2.3cm} p{2.2cm} p{2.0cm} p{3cm}}
  \hline
 Process                                      & BSE-CIA&Expt.\cite{belle04}   &NRQCD\cite{braaten03}& \cite{luchinsky}&\cite{braguta}\\
  \hline
  $e^- e^+\rightarrow J/\psi(1S) J/\psi(1S)$  & 2.4549&$< 9.1$    &6.65$\pm$3.02   &2.260& 2.12$\pm$0.85 \\
  $e^- e^+\rightarrow \psi(2S) \psi(2S)$ & 1.6281 & $<5.2$         &1.15$\pm$0.52  &  0.230 & 0.24$\pm$0.10   \\
     \hline
  \end{tabular}
\caption{Cross sections (in fb) at leading order (LO) for process, $e^- e^+\rightarrow J/\Psi J/\Psi$ calculated in present work at $\sqrt{s}=10.6 GeV$ along with results of other models}
\end{center}
\end{table}

\begin{table}[hhhhh]
  \begin{center}
  \begin{tabular}{p{6.3cm} p{2.2cm} p{2.2cm} p{2.2cm} }
  \hline
 Process                                      & $N_V$&  $G_1$ & $G'_1$\\
  \hline
  $e^- e^+\rightarrow J/\psi(1S) J/\psi(1S)$  & 7.2958   & 0.00826      &0.00668  \\
  $e^- e^+\rightarrow \psi(2S) \psi(2S)$      & 5.9281    &0.00882      & 0.00729     \\
     \hline
  \end{tabular}
\caption{Numerical values of BS normalizer $N_V$ (in $GeV^{-2})$, and the radial integrals  $G_1$ (in $GeV^3$) , and $G'_1$ (in $GeV^4$)}
\end{center}
\end{table}

\section{Cross section for $\eta_c + \eta_c$ production in $e^- e^+$ annihilation}
The process $e^- e^+\rightarrow \gamma^*\gamma^*\rightarrow \eta_c +\eta_c$ takes place only through quark rearrangement (non-fragmentation) diagrams in Fig. 1, due to the fact that each of the $\eta_c$ has $C=+1$, and hence the fragmentation process is ruled out. The amplitude for the first rearrangement diagram can be written as,

\begin{equation}
{M^1}_{fi}|_D=\frac{16}{s^2}e^2 e_{Q}^{2}[\bar{v}^{(s2)}(\bar{p_2})\gamma_{\nu}[\frac{-i{\not}Q+m_e}{Q^2+m_e^2}]\gamma_{\mu}u^{(s1)}(\bar{p_1})]\int \frac{d^4q}{(2\pi)^4} \int \frac{d^4q}{(2\pi)^4}Tr[\gamma_{\mu}\bar{\Psi}(P,q)\gamma_{\nu}\bar{\Psi}(P',q')]
\end{equation}

The amplitude for the other fragmentation diagram that involves exchange of final state mesons for this process can again be obtained from the above amplitude through the exchange of indices as, $M^2_{fi}|_X=M^2_{fi}|_{D}~{\mu \leftrightarrow \nu}$.

To reduce the above amplitudes to 3D form, we make use of the Covariant Instantaneous Ansatz (CIA), where the 4D volume elements of internal momenta of the two produced hadrons can be written as $d^4q=d^3\hat{q} Md\sigma$ (and $d^4q'=d^3\hat{q}' M'd\sigma'$). We then integrate over the longitudinal components $Md\sigma$, and $M'd\sigma'$ of internal momenta $q$ and $q'$ of the two hadrons to obtain,$\int \frac{Md\sigma}{2\pi i}\hat{\Psi}(P,q)=\bar{\Psi}(\hat{q})$, and $\int \frac{M'd\sigma'}{2\pi i}\hat{\Psi}(P',q')=\bar{\Psi}(\hat{q}')$.

Thus, for the Direct amplitude of photon fragmentation diagrams, we reduce the amplitude to 3D form as,
\begin{equation}
{M^1}_{fi}|_D=\frac{16e^2e_Q^2}{s^2}[\bar{v}^{(s2)}(\bar{p_2})\gamma_{\nu}[\frac{-i{\not}Q+m_e}{Q^2+m_e^2}]\gamma_{\mu}u^{(s1)}(\bar{p_1})]\int \frac{d^3\hat{q}}{(2\pi)^3} \int \frac{d^3\hat{q}'}{(2\pi)^3} Tr[\gamma_{\mu}\bar{\Psi}(\hat{q})\gamma_{\nu}\bar{\Psi}(\hat{q}')]
\end{equation}

The amplitude for the exchange diagram for this process can be then be similarly written as,

\begin{equation}
{M^1}_{fi}|_X=\frac{16e^2e_Q^2}{s^2}[\bar{v}^{(s2)}(\bar{p_2})\gamma_{\nu}[\frac{-i{\not}Q+m_e}{Q^2+m_e^2}]\gamma_{\mu}u^{(s1)}(\bar{p_1})]\int \frac{d^3\hat{q}}{(2\pi)^3}\int \frac{d^3\hat{q}'}{(2\pi)^3} Tr[\gamma_{\nu}\bar{\Psi}(\hat{q})\gamma_{\mu}\bar{\Psi}(\hat{q}')].
\end{equation}

For pseudoscalar mesons, the adjoint wave functions after reduction to 3D form can be written as\cite{eshete19},
\begin{eqnarray}
&&\nonumber \bar{\Psi}_P(\hat{q})=N_P\gamma_5[M-i{\not}P+\frac{2{\not}q{\not}P}{M}]\phi_P(\hat{q})\\&&
\bar{\Psi}_P(\hat{q}')=N_P\gamma_5[M'-i{\not}P'+\frac{2{\not}q'{\not}P'}{M'}]\phi_P(\hat{q}'),
\end{eqnarray}

where the structure of radial wave functions $\phi_P(\hat{q})$ is the same as $\phi_V(\hat{q})$ in Eq.(16), but with $\beta_V \rightarrow \beta_P$ (for details see \cite{eshete19}). The total amplitude after trace evaluation over the gamma matrices can be written as,
\begin{eqnarray}
&&\nonumber M_{fi}=2\frac{16e^2e_Q^2}{s^2}\int \frac{d^3\hat{q}}{(2\pi)^3}\phi_P(\hat{q})\int \frac{d^3\hat{q}'}{(2\pi)^3}\phi_P(\hat{q}')\\&&
\nonumber \bigg[(-4MM'+4P.P')[\bar{v}^{(s2)}(\bar{p_2})[\frac{i{\not}Q+4m_e}{Q^2+m_e^2}]u^{(s1)}(\bar{p_1})]+\\&&
\nonumber ~~(-4-\frac{16\hat{q}.\hat{q}'}{MM'})[\bar{v}^{(s2)}(\bar{p_2}){\not}P\frac{-i{\not}Q+m_e}{Q^2+m_{e}^2}{\not}P'u^{(s1)}(\bar{p_1})+\bar{v}^{(s2)}(\bar{p_2}){\not}P'\frac{-i{\not}Q+m_e}{Q^2+m_{e}^2}{\not}Pu^{(s1)}(\bar{p_1})]-\\&&
~~~~\frac{16P.P'}{MM'}[\bar{v}^{(s2)}(\bar{p_2}){\not}\hat{q}\frac{-i{\not}Q+m_e}{Q^2+m_{e}^2}{\not}\hat{q}'u^{(s1)}(\bar{p_1})+\bar{v}^{(s2)}(\bar{p_2}){\not}\hat{q}'\frac{-i{\not}Q+m_e}{Q^2+m_{e}^2}{\not}\hat{q}u^{(s1)}(\bar{p_1})]\bigg]
\end{eqnarray}

Taking $I_{\mu}$ as the unit vector along the direction of $\hat{q}_{\mu}$, introducing the 3D integrals $G_1, G_2, G'_1, G'_2$ defined in Eq.(19), and making use of the fact that in lines 3 and 4 of Eq.(34), the second term on the right contributes equally as the first term on the left, the amplitude simplifies as,

\begin{eqnarray}
&&\nonumber M_{fi}=\frac{16e^2e_Q^2}{s^2}\bigg[2G_1 G_2(-4MM'+4P.P')[\bar{v}^{(s2)}(\bar{p_2})[\frac{i{\not}Q+4m_e}{Q^2+m_e^2}]u^{(s1)}(\bar{p_1})]+\\&&
4(-4G_1G_2-\frac{16I.I'}{MM'}G'_1G'_2)[\bar{v}^{(s2)}(\bar{p_2}){\not}P\frac{-i{\not}Q+m_e}{Q^2+m_{e}^2}{\not}P'u^{(s1)}(\bar{p_1})]-
\frac{16P.P'}{MM'}G'_1G'_2[\bar{v}^{(s2)}(\bar{p_2}){\not}\hat{q}\frac{-i{\not}Q+m_e}{Q^2+m_{e}^2}{\not}\hat{q}'u^{(s1)}(\bar{p_1})\bigg].
\end{eqnarray}

Here in the photon propagators, we have used, $k^2=-\frac{s}{4}$. We now evaluate the spin averaged amplitude square given as, $|\bar{M}_{fi}|^2=\frac{1}{4}\sum_{s_1, s_2}M^{\dag}_{fi}M_{fi}$, and can be expressed as:

\begin{eqnarray}
&&\nonumber |\bar{M}_{fi}|^2= -\frac{2^{24}\pi^4\alpha^4}{3^4M^2M'^2s^4(m_{e}^2+\frac{s}{4})^2}{N_{P}}^2 {N'_{P}}^2\bigg[\frac{1}{2}G_{1}^2G_{2}^2X_1-8G_1G_2MM'(-4G'_1G'_2+G_1G_2MM')X_2+\\&&
\nonumber ~~~~~~~~~~~~~~~~~~~~~~~~~~~~~~~~~~~~~~~~~~~~~~~8(-4{G'}_1 {G'}_2+G_{1}G_{2} 2MM')^2X_3+64{G'_{1}}^2{G'_{2}}^2 X_4+16G'_1G'_2(4{G'}_1{G'}_2 {G}_1{G}_2 MM')X_5\bigg];\\&&
\nonumber X_1=M^2M'^2(-2M^2+2MM'+s)^2(-2m_{e}^2s+s^2),\\&&
\nonumber X_2=s^3(2M^2-2MM'-s)(1+2Cos\theta+Cos2\theta),\\&&
\nonumber X_3=32M^2M'^2m_{e}^4-8M^2M'^2m_{e}^2s-M'^2s^3(-1+Cos\theta)(1+3Cos\theta)+8m_{e}^2M'^2s(2M^2+sCos\theta(1-Cos\Theta)+\\&&
\nonumber ~~~~8M^2m_{e}^2s(2M'^2+sCos\theta(1-Cos\theta)+s^2(-1+Cos\theta)^2(2M^2s-s^2Sin^2\theta)-\\&&
\nonumber ~~~~2m_{e}^2s[8M^2M'^2-7M^2s-M'^2s+3s^2+4s(2M^2-s)Cos\theta+ s(-M^2+M'^2+s)Cos2\theta],\\&&
\nonumber X_4=(-2M^2+s^2)^2s^2(-1+Cos4\theta),\\&&
X_5=(2M^2-s)[-3s^3-6s^3Cos\theta-4s^3Cos2\theta-2s^3Cos3\theta-s^3Cos4\theta.
\end{eqnarray}

Here $\alpha=\frac{e^2}{4\pi}$ is the QED coupling constant, while, $N_P$ and $N'_P$ are the BS normalizers of pseudoscalar mesons and are evaluated through the current conservation condition, Eq.(27). The total cross section for the process is given as,

\begin{equation}
\sigma=\frac{1}{32\pi^2 s^{3/2}}|\vec{P'}|\int d\Omega' |\bar{M}_{fi}|^2,
\end{equation}

with $|\vec{P'}|=\sqrt{\frac{1}{s}[s-(M+M')^2][s-(M-M')^2]}$ the momenta of either of the ejecting particles. The cross section values for $e^- e^+\rightarrow \eta_c + \eta_c$ is given in Table 2.

\begin{table}[hhhhh]
  \begin{center}
  \begin{tabular}{p{6.3cm} p{3.1cm} p{3.5cm} }
  \hline
 Process                                      & BSE-CIA& NRQCD\cite{braaten03}\\
  \hline
  $e^- e^+\rightarrow \eta_c(1S) \eta_c(1S)$  & $2.087 \times 10^{-3}$ & $(1.83\pm 0.10)\times 10^{-3}$ \\
  $e^- e^+\rightarrow \eta_c(2S) \eta_c(2S)$ & $0.2369\times 10^{-3}$ & $(0.31\pm 0.02)\times 10^{-3}$  \\                                             \\
     \hline
  \end{tabular}
\caption{Cross sections (in fb) at leading order (LO) for process, $e^- e^+\rightarrow \eta_c \eta_c$ calculated in present work at $\sqrt{s}=10.6 GeV$ along with results of other models}
\end{center}
\end{table}

\begin{table}[hhhhh]
  \begin{center}
  \begin{tabular}{p{6.3cm} p{2.2cm} p{2.2cm} p{2.2cm} }
  \hline
 Process                                      & $N_P$& $ G_1$& $G'_1$\\
  \hline
  $e^- e^+\rightarrow \eta_c(1S) \eta_c(1S)$  & 7.6452   & 0.008119      &0.006527  \\
  $e^- e^+\rightarrow \eta_c(2S) \eta_c(2S)$     & 7.38021    &-0.00877      & -0.01206     \\
     \hline
  \end{tabular}
\caption{Numerical values of BS normalizer $N_P$ (in $GeV^{-2})$, and the radial integrals  $G_1$ (in $GeV^3$) , and $G'_1$ (in $GeV^4$)}
\end{center}
\end{table}

\section{Conclusion}
We have studied the processes $e^- e^+\rightarrow J/\psi+J/\psi$, and $e^- e^+\rightarrow \eta_c+\eta_c$ in the framework of $4\times 4$ Bethe-Salpeter equation. The double $J/\psi$ production process is dominated by photon fragmentation diagrams. The total cross section for this process is $\sigma(e^- e^+\rightarrow J/\psi J/\psi)$=2.4549fb, which is within the experimental upper bound, $\sigma <$ 9.2fb\cite{belle04}. The contribution of fragmentation diagrams is $\sigma^{frag.}=$3.21574fb (which is 131$\%$ of total cross section) while the contribution from the quark rearrangement diagrams is $\sigma^{Rearr.}$=0.3639fb. Thus the contribution to cross section from the interference between the two sets of diagrams is $\sigma^{Int.}$=-1.1247fb. Our result for $\sigma(e^- e^+\rightarrow \Psi(2S) \Psi(2S))$=1.6281fb is again within the experimental upper bound $\sigma <$ 5.2 fb \cite{belle04} for this processes, as can be seen from Table 1, where we have also compared our results with other models.

We have also calculated the cross section for double $\eta_c$ production, which receives contribution only from the quark rearrangement diagrams. We have obtained the cross section for this process as, $\sigma(e^- e^+\rightarrow \eta_c \eta_c)=2.087\times 10^{-3}$fb, whose cross section is nearly three orders of magnitude smaller than $\sigma(e^- e^+\rightarrow J/\psi J/\psi)$, and have compared our result with NRQCD\cite{braaten03}. The experimental data for this process is not yet available.\\

\bigskip
Acknowledgement: One of us (SB) would like to thank Theoretical Physics Department, CERN for his visit during May-June 2024, where a major part of this work was performed. He would like to thank Andreas Juttner for valuable discussions.

\end{document}